\documentclass[reprint,amsmath,amssymb,aps,twocolumn,nofootinbib]{revtex4-2}
\usepackage{graphicx,amsmath,amsfonts,bm, color,MnSymbol}

\usepackage{bm}
\usepackage{xcolor,hyperref}
\hypersetup{
   colorlinks,
   linkcolor={blue!50!black},
   citecolor={blue!50!black},
   urlcolor={blue!80!black}
}
\newcommand*\colvec[3][]{\begin{pmatrix}\ifx\relax#1\relax\else#1\\\fi#2\\#3\end{pmatrix}}
\newcommand*\rowvec[3][]{\begin{pmatrix}\ifx\relax#1\relax\else#1&\fi#2&#3\end{pmatrix}}

\newcommand{\Grad}{\nabla}

\newcommand{\Div}{\nabla\cdot}

\newcommand{\mycomment}[1]{}

\newcommand{\UCD}[1]
{\stackrel{\smalltriangledown}{#1}}

\newcommand{\units}[1]{\left[#1\right]}

\begin{document}

\title{Nonlinear diffusion and compressive rims in source-driven biopolymer condensates}
\author{Avraham Moriel and Howard A. Stone}
\affiliation{Department of Mechanical and Aerospace Engineering, Princeton University, Princeton, New Jersey 08544, USA}

\begin{abstract}
    Many subcellular condensates continuously produce biopolymers. Coupling Flory-Huggins thermodynamics to two-fluid viscoelasticity, we probe the diffusion of such source-driven polymeric droplets, and identify a universal structural compressive rim at their diffusion front. Integrating analytical scaling laws, numerical simulations, and experimental data, we show that this framework captures key structural and dynamic characteristics of the nucleolus, demonstrating the role of polymer diffusion in non-equilibrium biological transport.
\end{abstract}
\maketitle

Polymer dynamics is crucial in both biological systems and myriad technological applications~\cite{de1979scaling,rubinstein2003polymer,doi1988theory}. Not surprisingly, polymer diffusion and interdiffusion have been longstanding research themes describing how entanglements and solvent quality govern polymeric transport, in particular, dynamically coupling macroscopic transport to microscopic chain dynamics~\cite{de1980dynamics,pincus1981dynamics,de1971reptation,doi1978dynamics1,doi1978dynamics2,doi1978dynamics3,brochard1983polymer,brochard1991kinetics,brochard1983polymer,brochard1991kinetics,green1987diffusion,jordan1988mutual,brochard1986polymer,kausch1989polymer,klein1990interdiffusion,kramer1984interdiffusion,green1985marker,kramer1984polymer,yetter1983flux,tead1992interdiffusion,mills1984analysis,yokoyama2000mutual}. This interplay between thermodynamics and structural relaxation is also central to processes within biological cells, where membraneless organelles have been understood as liquid-liquid phase-separated condensates~\cite{brangwynne2015polymer,hyman2014liquid,brangwynne2009germline,shin2017liquid,berry2018physical,zwicker2014centrosomes}. Such an interplay is captured by a two-phase viscoelastic formalism, accommodating complex rheological responses while accounting for the mutual friction between the polymer network and the solvent~\cite{milner1991hydrodynamics,doi1992dynamic,mavrantzas1992modeling,larson1992flow,beris1994thermodynamics,beris1994compatibility,tanaka1996universality,tanaka1997viscoelastic,tanaka2000viscoelastic,tanaka2012viscoelastic}. While such an approach successfully captures the formation of stable condensates and the emergence of gel-like morphologies~\cite{tanaka1993unusual,taniguchi1996network,tanaka1997phase,araki2001three,zhang2001kinetics,tanaka2006viscoelastic,doi2009gel,tanaka2022viscoelastic}, the mixing of polymer phases into their surroundings remains comparatively unexplored within biological context (polymer-polymer interdiffusion has been a long-standing problem~\cite{de1980dynamics,pincus1981dynamics,doi1992dynamic,brochard1983polymer,brochard1991kinetics,jordan1988mutual,green1987diffusion,brochard1986polymer,kausch1989polymer,klein1990interdiffusion,kramer1984interdiffusion,green1985marker,kramer1984polymer,yetter1983flux,tead1992interdiffusion,mills1984analysis,yokoyama2000mutual}).

Transport is central to function in many biophysical systems. Here we focus on the nucleolus --- a subnuclear condensate synthesizing and exporting ribosomal RNA (rRNA)~\cite{boisvert2007multifunctional,carmo2000or,riback2023viscoelasticity}. The nucleolus acts as a persistent source, driving an outward flux of biopolymers into the surrounding nucleoplasm. Recent experiments have characterized the nucleolus as a viscoelastic medium~\cite{riback2023viscoelasticity,cheng2025micropipette,feric2016coexisting}, suggesting that the resulting transport of rRNA follows advection-diffusion dynamics~\cite{riback2023viscoelasticity}. Structurally, the internal regions of the nucleolus are characterized by extended rRNA morphologies, while the rRNAs are compressed at the nucleolus's rim~\cite{feric2016coexisting,riback2023viscoelasticity,miller1981nucleolus,mougey1993terminal}. Our aim is to unite the structural and dynamic empirical observations, linking the thermodynamics of the polymer-solvent mixture to the observed flow-induced deformations. 

In this Letter, we couple Flory-Huggins thermodynamics to polymer viscoelasticity to study a class of nonlinear diffusion equations~\cite{brochard1983polymer,brochard1991kinetics,doi1992dynamic} driven by the polymer source. Generically, we find that the osmotic pressure self-induces a structural ``compressive rim'' at the diffusion front, implying compressed polymer conformations. This physical picture complements recent works showing that condensates can exert measurable mechanical stresses on their surroundings~\cite{style2018liquid,rosowski2020elastic,rosowski2020elastic}.  We study analytically and numerically how the temporal nature of the source affects the emerging dynamics. Finally, we utilize this framework to revisit recent experimental data involving the nucleolus, from which we estimate the nucleolar rRNA production rate time-dependence.

\emph{Osmotic-driven polymer diffusion.} --- We consider a two-component mixture, consisting of polymeric species A and B, referred to as A-mers and B-mers~\cite{milner1991hydrodynamics,doi1992dynamic,mavrantzas1992modeling,larson1992flow,beris1994thermodynamics,beris1994compatibility,tanaka1996universality,tanaka2000viscoelastic,tanaka2012viscoelastic,tanaka1997viscoelastic}. The A-mers' monomer volume fraction field $\phi_A(\bm{r},t)$ is driven by the presence of a localized source $S(\bm{r},t)$ representing a continuous synthesis of A-mers. The polymer flux from the source creates a concentrated region that diffuses outwards due to gradients in the osmotic pressure $\bm{\Pi}$. This transport is accompanied by flow-induced conformational deformations of the polymers and frictional resistance from the surrounding medium.

\begin{figure}[b]
\includegraphics[width=0.46\textwidth]{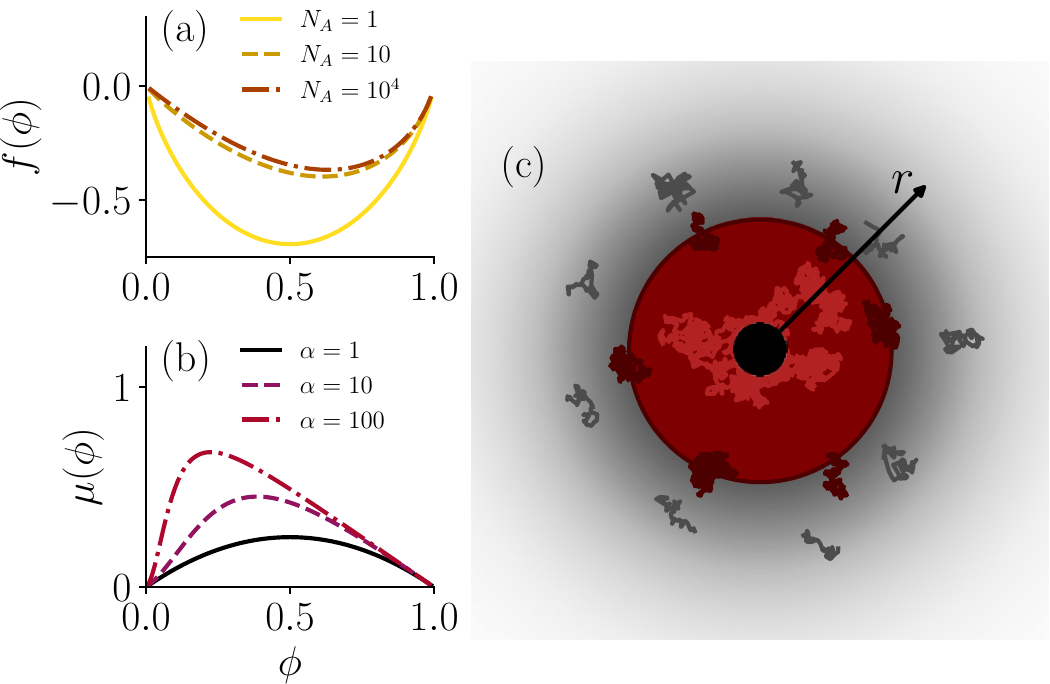}
\caption{Two component polymer mixtures. (a) The Flory-Huggins thermodynamic potential $f$, Eq.~\ref{eq:FH}, for non-interacting species ($\chi\!=\!0$), as a function of the A-mer volume fraction $\phi_A\!=\!\phi$, for different $N_A$ values (with $N_B\!=\!1$). (b) The friction function $\mu(\phi)$ for different values of the entanglement parameter $\alpha$. (c) Sketch of a source (black circle) of long A-mers (red) inducing A-mer diffusion (dark red circle) into a medium of short B-mers (gray). A-mers and B-mers are schematically plotted as red and gray random chains, respectively.}
\label{fig:fig1}
\end{figure}

We define the A-mer and B-mer velocity fields as $\bm{u}_A(\bm{r},t)$ and $\bm{u}_B(\bm{r},t)$, respectively 
[below we suppress the explicit $(\bm{r},t)$ dependence for readability]. The volume fractions $\phi_A$ and $\phi_B$ evolve via the continuity equations
\begin{equation}\label{eq:con_clean}
\begin{aligned}
    \partial_t \phi_A + \nabla\cdot\left(\bm{u}_A \phi_A\right) &= S \ , \\
    \partial_t \phi_B + \nabla\cdot\left(\bm{u}_B \phi_B\right) &= 0,
\end{aligned}
\end{equation}
and because we consider a binary mixture, $\phi_A + \phi_B\!=\!1$. For simplicity, we denote $\phi_A\!=\!\phi$.

The pressure $p$, osmotic pressure $\bm{\Pi}$, A-stresses $\bm{\sigma}_{A}$, and B-stresses $\bm{\sigma}_{B}$ contribute to the quasi-static momentum balance~\cite{tanaka2000viscoelastic,tanaka2012viscoelastic},
\begin{equation}
    {\bm 0} = \nabla\cdot\left(-p\bm{I}-\bm{\Pi} + \boldsymbol{\sigma}_{A} + \boldsymbol{\sigma}_{B}\right)  \ , \label{eq:qs_clean}
\end{equation}
where $\bm{I}$ is the identity tensor.

The relative motion of the two species is obtained as~\cite{doi1992dynamic,tanaka2000viscoelastic,beris1994thermodynamics}
\begin{equation}\label{eq:rel_clean}
    \bm{u}_A - \bm{u}_B = -\frac{1}{ n_0\zeta_0\mu}\left[\left(1-\phi\right)\nabla\cdot\left(\bm{\Pi}-\bm{\sigma}_{A}\right)+\phi\nabla\cdot\bm{\sigma}_{B}\right] \ ,
\end{equation}
where $n_0$ is the mixture's number density (monomers per volume), $\zeta_0$ is a monomer friction coefficient, and $\mu\!\equiv\!\mu\left(\phi\right)$ is a dimensionless friction function that depends on the polymers and their surroundings~\cite{brochard1983polymer,brochard1991kinetics,doi1992dynamic}. Given the source $S$, the friction function $\mu$, the osmotic pressure $\bm{\Pi}$, and the stresses $\bm{\sigma}_A$ and $\bm{\sigma}_B$, Eqs.~\eqref{eq:con_clean}-\eqref{eq:rel_clean} determine $\phi$, $p$, $\bm{u}_A$, and $\bm{u}_B$.

The osmotic pressure originates from polymer concentration, i.e. chemical potential, gradients. We utilize the Flory-Huggins (FH) thermodynamic potential~\cite{flory1953principles,huggins1942theory} $F_{FH}\!=\!k_B T n_0 f(\phi)$, where $k_B$ is the Boltzmann constant, $T$ is the temperature, and
\begin{equation}\label{eq:FH}
    f(\phi) = \frac{\phi}{N_A} \log \phi + \frac{\left(1-\phi\right)}{N_B} \log\left(1-\phi\right) + \chi \phi \left(1-\phi\right) \ .
\end{equation}
Here, $N_A$ and $N_B$ are the number of monomers per A-mer and B-mer respectively, and $\chi$ is an energy interaction parameter between the two species (we omit surface tension); $f(\phi)$ is plotted for non-interacting polymers  considered in what follows ($\chi\!=\!0$), for various $N_A$ values (and where $N_B\!=\!1$) in Fig.~\ref{fig:fig1}(a). For the FH potential of Eq.~\eqref{eq:FH}, the resulting osmotic pressure is isotropic, obtained as $\bm{\Pi}\!=\!k_B T n_0\bm{I} \left(\phi\partial_\phi f-f\right)$~\cite{SM,LongPaper}.

The stresses $\bm{\sigma}_{A}$ and $\bm{\sigma}_{B}$ evolve via an upper-convected Maxwell (UCM) model~\cite{tanaka2000viscoelastic}, 
\begin{equation} \label{eq:UCM}
    \UCD{\bm{\sigma}}_{(i)} = G_{(i)}\left(\bm{L}_{(i)} + \bm{L}_{(i)}^T\right) - \frac{1}{\lambda_{(i)}}\bm{\sigma}_{(i)} \ ,
\end{equation}
where $i$ denotes either A or B (no summation implied), $\UCD{\bullet}\equiv\partial_t \bullet + \bm{u}_{(i)} \cdot \nabla \bullet -\left[\bm{L}_{(i)}^T\cdot \bullet + \bullet \cdot \bm{L}_{(i)}\right]$ is the upper-convected time derivative, $G_{(i)}$ is the shear modulus, $\lambda_{(i)}$ is the relaxation time, and $\bm{L}_{(i)}\!\equiv\!\nabla \bm{u}_{(i)}$ is the velocity gradient tensor for the $i$th component.

\begin{figure}[b]
\includegraphics[width=0.48\textwidth]{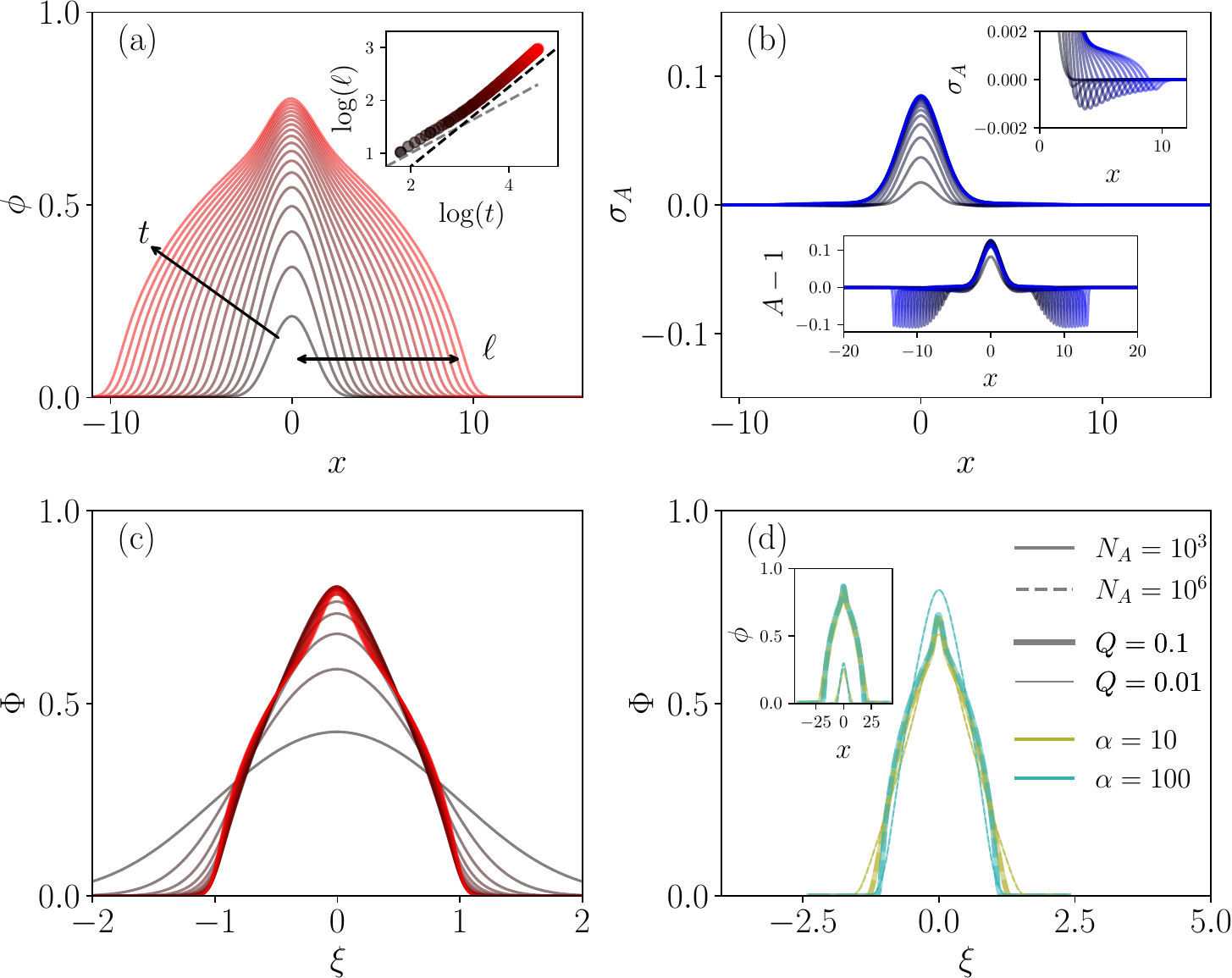}
\caption{Dynamics of a 1D diffusing droplet. (a) The numerical solution (with $N_A\!=\!10^4$, $N_B\!=\!1$, $\alpha\!=\!100$, $\beta\!=\!1$, $\tilde{G}\!=\!1$, and $Q\!=\!0.1$) for the volume fraction $\phi$ at increasing times (gray for early times to red at later times), demonstrating the source-driven polymer droplet diffusing into the solvent. Inset: log-log plot of the characteristic droplet radius $\ell$ as a function of time, showing a transition from linear to nonlinear diffusion. The dashed lines are guides-to-the-eye for linear diffusion $\ell\sim t^{1/2}$ (in gray), and nonlinear diffusion $\ell\sim t^{3/4}$ (in black). (b) The numerically-obtained 1D stress $\sigma_A$ [using similar parameters as in (a)] at different times (gray for early times to blue at later times). Top inset: a zoom-in on the compressive region at the edge of the expanding polymer droplet. Bottom inset: the associated polymer conformation $A-1\!=\!\sigma_A/\phi$, showing compressed conformations at the expanding edges. (c) The rescaled polymer volume fraction $\Phi(\xi)\!=\!\phi / \left(Q^2 t\right)^{1/4}$ from (a), as a function of the rescaled coordinate $\xi\!\equiv\!x/(Q^2 t^3)^{1/4}$ at different times (gray for early times to red at later
times). (d) Eight rescaled $\Phi$ profiles obtained for various combinations of $N_A$, $Q$, and $\alpha$ values (see legend), at $t\!=\!200$. Inset: $\phi$ versus $x$ solutions, demonstrating $Q$ controls the dynamics ($\alpha$ and $N_A$ barely change them).}
\label{fig:fig2}
\end{figure}

Finally, the polymeric friction $\mu\!=\!\frac{\mu_A \mu_B}{\mu_A + \mu_B}$ depends on the individual friction functions $\mu_A(\phi)$ and $\mu_B(\phi)$ via their degree of entanglements~\cite{brochard1983polymer,brochard1991kinetics,doi1992dynamic}. Unentangled, short polymers correspond to  $\mu_{(i)}\!=\!\phi_{(i)}$, while entangled polymers have a steeper slope $\mu_{(i)}\!=\!\alpha_{{(i)}} \phi_{(i)}$ (without summation), depending on the ratio between the polymer length $N_{(i)}$ and a typical entanglement threshold $N_e$, $\alpha_{(i)}\!=\!N_{(i)}/N_e$ (with $N_{(i)}\!\ge\!N_e$ so that $\alpha\!\ge\!1$)~\cite{LongPaper}. 

We consider long, slowly relaxing A-mers, and short, quickly relaxing B-mers, with $N_A\!\gg\!N_B$. For a sufficiently small $\lambda_B$, $\bm{\sigma}_B\!\simeq\!\eta_B\left(\bm{L}_B+\bm{L}_B^T\right)$ responds as a Newtonian fluid with effective viscosity $\eta_B\!\equiv\!G_B \lambda_B$, and the system evolves on the A-mer relaxation time $\lambda_A\!\equiv\!\lambda$. For simplicity, we consider $G_A\!\simeq\!\phi G_0$ to respond proportionally to the local volume fraction $\phi$ multiplied by an effective modulus $G_0$ (valid for low concentrations~\cite{beris1994thermodynamics,ferry1961viscoelastic,johnson1970infinite}; nonlinear $G_A(\phi)$ relations such as those suggested in~\cite{larson2015modeling,milner2005predicting,mackintosh1995elasticity} are not expected to qualitatively change what follows). Additionally, we take $N_B\!<\!N_e$, implying the B-mers are not entangled, hence $\mu_B\!=\!1-\phi$. To allow A-mers to transition from entangled to unentangled states, we choose $\mu_{A}\!=\!\phi\left[\alpha\phi+(1-\phi)\right]$~\cite{SM,LongPaper}, resulting in $\mu(\phi)$ as plotted in Fig.~\ref{fig:fig1}(b). Using these assumptions, we now study a source producing A-mers that diffuse into a medium of B-mers, schematically depicted in Fig.~\ref{fig:fig1}(c).

\emph{Nonlinear diffusion and self-induced compressive rim.} --- First, we consider an A-mer droplet of typical radius $\ell$ diffusing into a  solution of B-mers in the absence of a source $S$. For negligible material stresses $\bm{\sigma}_{A}$ and $\bm{\sigma}_{B}$, we sum Eq.~\eqref{eq:con_clean}, and combine this relation with Eq.~\eqref{eq:rel_clean} to obtain the velocity field $\bm{u}_A$~\cite{LongPaper}. Substituting into Eq.~\eqref{eq:con_clean} yields a nonlinear diffusion equation~\cite{brochard1983polymer,brochard1991kinetics,doi1992dynamic,LongPaper}
 \begin{equation}\label{eq:nonlin}
    \partial_t \phi \simeq \nabla\cdot\left[\frac{\left(1-\phi\right)^2\phi}{n_0 \zeta_0 \mu(\phi)}\nabla\cdot\bm{\Pi}\right] \ .
\end{equation}

At the origin $\bm{u}_A$ vanishes due to isotropy. The osmotic pressure drives the A-mers into the B-mer environment, implying that within the droplet, $r\!<\!\ell$, the velocity is directed outwards, $\bm{u}_{A}\cdot\hat{\bm r}\!>\!0$. Finally, outside the droplet, for $r\!\gg\!\ell$, the A-mers are static, $\bm{u}_A\!=\!{\bm 0}$. The resulting velocity profile necessitates a transition from an extensile region within the droplet $\bm{L}_A^{rr}\!>\!0$ to a compressive region $\bm{L}_A^{rr}\!<\!0$ at the expanding front. The local strain rates, proportional to $\bm{L}_A+\bm{L}_A^T$, induce material stresses $\bm{\sigma}_A$ according to Eq.~\eqref{eq:UCM}, leading to a region of compressive stresses at the droplet's boundary. Considering a source $S$, we expect a similar compressive rim at the diffusing front. 


\emph{Source-driven expansion in one dimension.} --- We first study the source-driven dynamics of a one-dimensional (1D) equivalent of Eqs.~\eqref{eq:con_clean}-\eqref{eq:UCM}~\cite{LongPaper}. For simplicity, we set $N_B\!=\!1$, and use $\lambda$ as a time scale, $\sqrt{\frac{k_B T \lambda}{\zeta_0}}$ as a length scale, and $n_0 k_B T$ as an energy density scale. These give rise to the dimensionless ratios $\beta\!\equiv\!\eta_B/n_0 k_B T \lambda$ and $\tilde{G}\!\equiv\!G_0/n_0 k_B T$, between the effective B-mer modulus $\eta_B/\lambda$ and elastic modulus $G_0$, and the thermal energy density $n_0 k_B T$, respectively~\cite{SM}.

We numerically solve~\cite{SM,van1995python,harris2020array} the 1D equivalent of Eqs.~\eqref{eq:con_clean}-\eqref{eq:UCM}, and show the evolution for the 1D volume fraction field $\phi$ in Fig.~\ref{fig:fig2}(a), subjected to a localized, constant Gaussian source $S\!=\!\frac{Q}{\sqrt{2\pi}}\exp{\left(-x^2/2\right)}$. At the initial phase, the source generates A-mers that linearly diffuse into the B-mers phase. As the polymer volume fraction increases, entropic forces push A-mers to diffuse into the B-mers phase, rendering the diffusive process nonlinear, and causing the droplet to expand with a temporal power-law different from the linear diffusion. The temporal evolution of a typical droplet radius $\ell$ (defined here as half the region where $\phi\!\ge\!0.1$), plotted in the inset of Fig.~\ref{fig:fig1}(a), shows a transition from $\ell\!\sim\! t^{1/2}$ to
$\ell\!\sim\!t^{3/4}$ diffusive dynamics~\cite{LongPaper}.

Next, we report the evolution of the stress $\sigma_A$ in Fig.~\ref{fig:fig2}(b), revealing $\sigma_A$ has an extensile peak around the source, and compressive stresses near the edges of the expanding droplet (top inset in panel b). Finally, many rheological models~\cite{beris1994thermodynamics,boyko2024perspective} developed for low-concentration polymeric fluids adopt the usage of a conformation tensor $\bm{A}$. In many such models $\bm{\sigma}_A\!\sim\!G_R n \left(\bm{A}-\bm{I}\right)$, where $G_R$ is a rheological modulus, $n$ is the polymer density, and $\bm{A}-\bm{I}$ is the deviation of the conformation tensor $\bm{A}$ from isotropic conformation~\cite{beris1994thermodynamics,boyko2024perspective}. We plot the equivalent quantity for our 1D model, $\sigma_A/\phi\!\propto\!A-1$, in the bottom inset of Fig.~\ref{fig:fig2}(b), implying the A-mers' conformations are compressed $A-1\!<\!0$ at the rim of the droplet.


Using the framework of Eq.~\eqref{eq:nonlin}, we anticipate a 1D, scaling solution $\phi(x,t)\!=\!\left(Q^2 t \right)^{1/4}\Phi(\xi)$, with $\xi\!=\!x / \left(Q^2 t^3\right)^{1/4}$~\cite{barenblatt2003scaling,stone2002partial,stone2025thin,pattle1959diffusion,van1977class,LongPaper}; note that the presence of a source introduces an additional flux term in Eq.~\eqref{eq:nonlin}, which respects the same scaling~\cite{LongPaper}. Rescaling the numerical results accordingly, the $\phi$ solutions collapse, as shown in Fig.~\ref{fig:fig2}(c). We also verify the generality of the scaling relations and volume fraction profiles for various combinations of A-mers' length $N_A$, source strength $Q$, and entanglement factor $\alpha$. We show eight rescaled volume fraction profiles obtained for various combinations in Fig.~\ref{fig:fig2}(d), revealing variations in the source amplitude $Q$ change the dynamics, while variations in $\alpha$ and $N_A$ leave the dynamics almost unchanged. All solutions adhere to the scaling relations for $\phi$ and $x$.


\begin{figure}[t]
\includegraphics[width=0.48\textwidth]{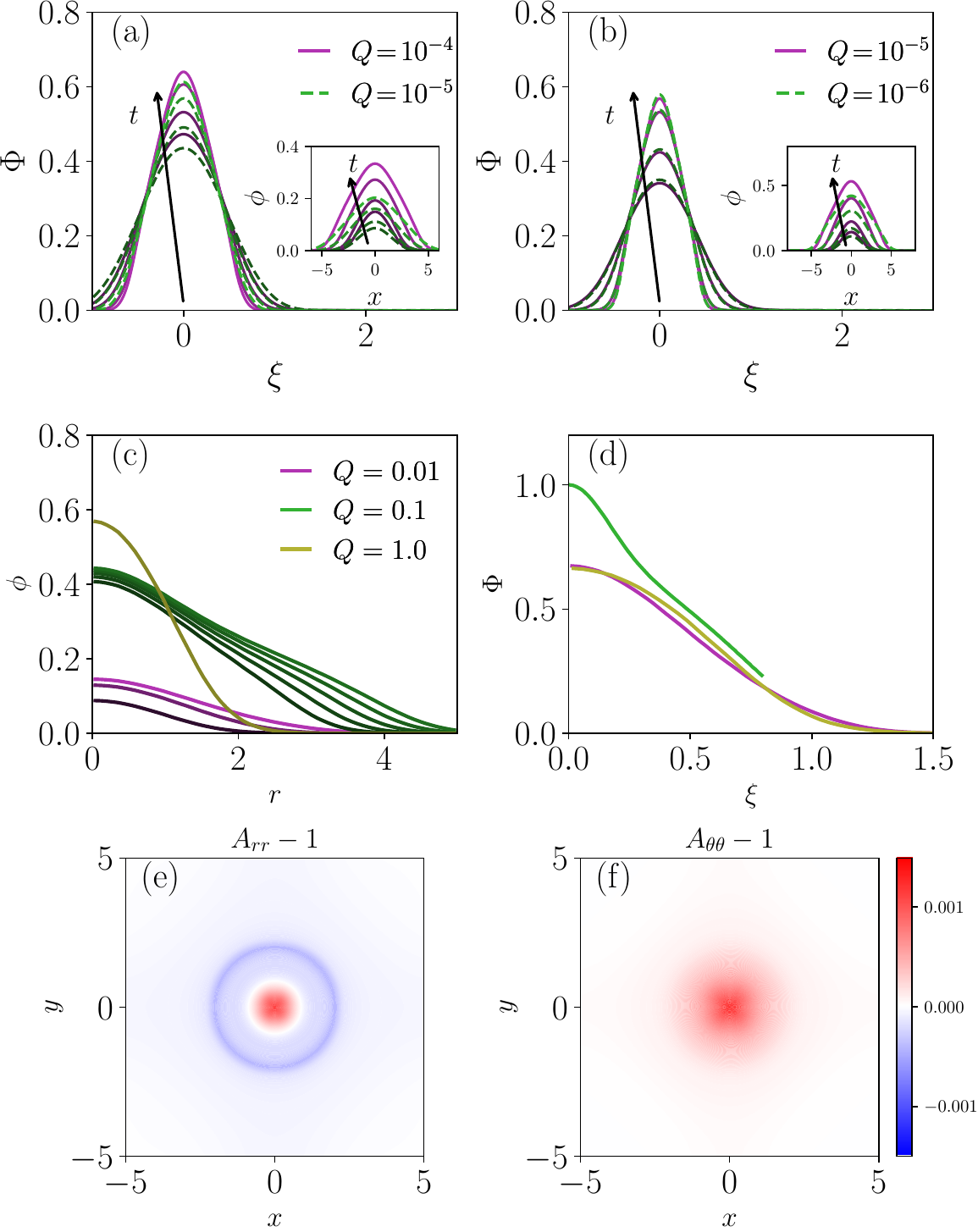}
\caption{Generalization of the scaling relations and the compressive rim. Self-similar solutions for (a) $q\!=\!2$ and (b) $q\!=\!3$, for two different $Q$ values ($N_A\!=\!10^4$, $N_B\!=\!1$, $\alpha\!=\!100$, $\beta\!=\!1$ and $\tilde{G}\!=\!1$), 
Insets: the corresponding raw $\phi$ fields. (c) Angular-averaged solutions $\phi(r)$ from the 2D numerical simulations with $q\!=\!1$ for $Q\!=\!0.01,0.1$ and $1.0$. (d) Collapse using $\phi\!=\!Q^{1/3} \Phi\left(\xi\right)$ with $\xi\!=\!r/\left(Q^2t^3\right)^{1/6}$. Snapshots of the radial and angular conformations $A_{rr}-1\!=\!\sigma_A^{rr}/\phi$, and $A_{\theta\theta}-1\!=\!\sigma_A^{\theta\theta}/\phi$ in (e) and (f) showing the extensile core and compressive radial rim in 2D.}
\label{fig:fig3}
\end{figure}

\emph{Time-variation, spatial dimensionality, and biological applications.} --- The formalism above offers a framework to understand biological processes~\cite{riback2023viscoelasticity}. Biological polymer sources need not be constant in time, and usually exist in higher spatial dimensions, implying that Eqs.~\eqref{eq:con_clean}-\eqref{eq:UCM} should be solved with the appropriate time variation of the source and in the proper spatial dimensionality.

To generalize our formalism~\cite{SM,LongPaper}, we revisit the nonlinear diffusion relation Eq.~\eqref{eq:nonlin}, together with the increase of total fraction $\int \phi dV\!=\!\frac{1}{q}Q t^{q}$, where $dV$ is an infinitesimal volume in $d$ spatial dimensions~\cite{SM}. Approximating Eq.~\eqref{eq:nonlin} as $\partial_t \phi \!=\!\nabla\cdot\left[D(\phi)\nabla\phi\right]$ with $D(\phi)\!=\!\phi^{m-1}$~\cite{SM,LongPaper}, and accounting for the different spatial scaling for $dV$, we obtain $\phi\!\simeq\!Q^{\gamma_1}t^{\delta_1}\Phi(\xi)$ with $\gamma_1\!=\!2/p$, $\delta_1\!=\!\left(2q-d\right)/p$, and $p\!\equiv\!2+d(m-1)$, and $r\!\simeq\! \ell \xi$ with $\ell\!\simeq\!Q^{\gamma_2} t^{\delta_2}$ with $\gamma_2\!=\!\left(m-1\right)/p$ and $\delta_2\!=\!\left[1+q(m-1)\right]/p$~\cite{SM,LongPaper}.

We show in Fig.~\ref{fig:fig3}(a)-(b) numerical 1D solutions for $\phi$ for different $q$ values representative of the source time-variability, and the application of the scaling relations above. Also, we solve a two-dimensional (2D; $d\!=\!2$) version of Eqs.~\eqref{eq:con_clean}-\eqref{eq:UCM}, and show in Fig.~\ref{fig:fig3}(c)-(d) the angular-averaged $\phi$ solutions and their scaled counterparts. Finally, we show in Fig.~\ref{fig:fig3}(e)-(f) snapshots of the radial and azimuthal 2D conformations, respectively as $A_{rr}-1$ and $A_{\theta\theta}-1$, which reveal the internal expansion and the compressive rim. In all cases, a compressive rim is present; and the scaling ansatz captures the dynamics.

We now apply this framework to the transport dynamics of the nucleolus, which is a subnuclear organelle responsible for synthesizing rRNA. Recent experiments revealed the nucleolus behaves as an entangled viscoelastic gel at its center, with a concentration that dilutes toward the periphery~\cite{riback2023viscoelasticity,cheng2025micropipette,feric2016coexisting}. The measurements suggested the rRNAs are extended within the dense core and folded at the nucleolar boundary~\cite{riback2023viscoelasticity,miller1981nucleolus,mougey1993terminal}, reminiscent of the  structural compressive rim discussed above.

\begin{figure}[t!]
\includegraphics[width=0.48\textwidth]{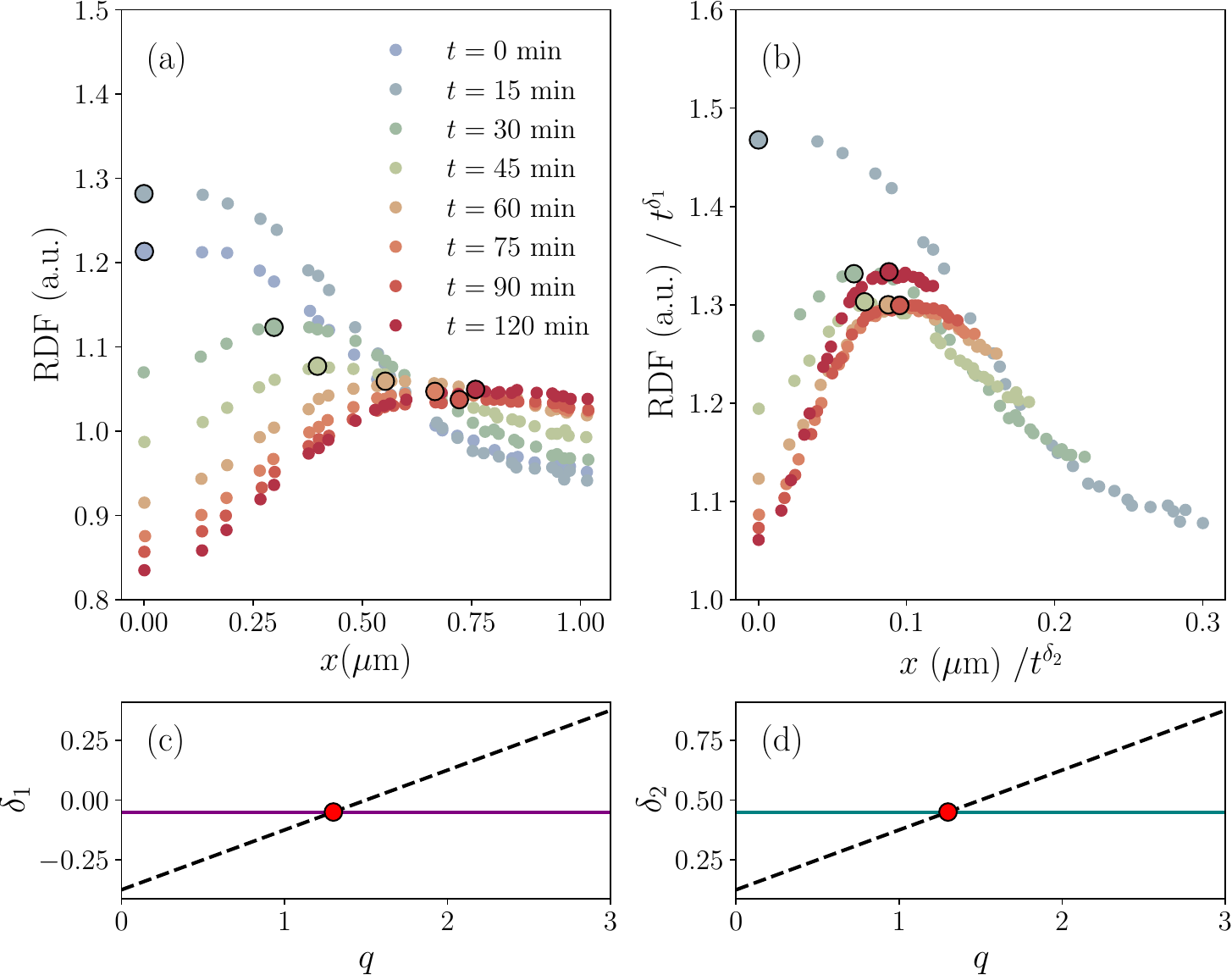}
\caption{Nucleolar expansion scaling. (a) RDFs at different times, digitized from Fig.4 (c) of~\cite{riback2023viscoelasticity}. Highlighted points show the maximal RDF for at different times. (b) The rescaled data $\text{RDF} / t^{\delta_1}$ versus $x/t^{\delta_2}$ revealing $\delta_1\!\simeq\!-0.05$ and $\delta_2\!\simeq\!0.45$. (c) The theoretical $\delta_1 \!=\!\left(2q-d\right)/p$ in dashed black line, and the observed value $-0.05$ in magenta, with an intersection at $q\!\simeq\!1.3$. (d) The theoretical $\delta_2\!=\!\left[1+q(m-1)\right]/p$ in dashed black, and the observed value $0.45$ in cyan, with an intersection at $q\!\simeq\!1.3$.}
\label{fig:fig4}
\end{figure}

To quantitatively test our theory, we analyze the experimental pulse-chase labeled data tracking the radial distribution function (RDF) of rRNA over time, digitized~\cite{WebPlotDigitizer} from Fig. 4(c) of~\cite{riback2023viscoelasticity}, as shown in Fig.~\ref{fig:fig4}(a). In the nonlinear diffusion framework, entangled rRNAs imply $N_A\!\gg\!N_e$, and $\alpha\!\gg\!1$, so $m\!\simeq\!3$~\cite{SM,LongPaper}. As the nucleolus is a three-dimensional organelle, we set $d\!=\!3$, leaving us with a single unknown parameter, the time-dependence of the rRNA production rate, $q$.

We rescale the spatial axis and RDF amplitude to collapse the curves, shown in Fig.~\ref{fig:fig4}(b), with $\delta_1\!\simeq\!-0.05$ and $\delta_2\!\simeq\!0.45$. Using $m\!=\!3$ and $d\!=\!3$, we plot $\delta_1(q)$ and $\delta_2(q)$ in Fig.~\ref{fig:fig4}(c)-(d) respectively, suggesting that $q\!\simeq\!1.3$. This implies that rate of production of rRNAs ($q-1$) lies in between a constant flux $q\!=\!1$, and a rate required to maintain a constant droplet boundary concentration $q\!=\!3/2$. Such an intermediate value reflects the non-equilibrium nature of the nucleolus, given the variability of its activity~\cite{iyer2018sirt7}.

\emph{Discussion. --- } We presented a theoretical framework for a source-driven diffusion of viscoelastic polymer droplets. Identifying the competition between osmotic pressure and polymer friction, we anticipated a structural compressive rim at the diffusion front, which we confirmed numerically. Experimentally, such a rim could potentially be verified using advanced imaging techniques sensitive to local structural anisotropy. Using scaling arguments, we derived the rescaling factors for the volume fraction $\phi$ and spatial coordinates ($x$ or $r$), depending on the spatial dimensionality $d$, the nonlinear diffusion exponent $m$, and the temporal exponent of the source $q$. Utilizing these relations resulted in the quantitative collapse of experimental rRNA profiles, suggesting the two-fluid polymer diffusion framework captures the non-equilibrium dynamics of active organelles like the nucleolus. The scaling relations suggest more detailed mathematical, similarity-like solutions~\cite{LongPaper}.

While the mean-field description above neglected enthalpic, electrostatic, and complex chain-level interactions, the robust scaling observed in the experimental data highlights the utility of this framework. Such an approach complements recent atomistic viewpoints~\cite{
das2013conformations,farag2022condensates,bauer2024conformations,garcia2026molecular,wani2026non} and other experimental efforts~\cite{michieletto2022rheology,fisher2024viscoelasticity}, offering a continuum framework to distinguish between passive and active, non-equilibrium transport often found in soft matter and biological condensates.

\emph{Acknowledgments. --- } A.M. acknowledges support from the \href{https://doi.org/10.37717/2021-3362}{James S. McDonnell Foundation Postdoctoral Fellowship Award in Complex Systems}. We acknowledge support from the NSF Grant DMS/NIGMS 2245850 and from the Princeton Center for Complex Materials (PCCM), an NSF-supported Materials Research Science and Engineering Center under award DMR-2011750.
The authors also thank N. Wingreen and C. P. Brangwynne for helpful comments and discussions. 

%


\onecolumngrid
\newpage
\begin{center}
\textbf{\large Supplemental Materials for: \\ ``Nonlinear diffusion and compressive rims in source-driven biopolymer condensates''}
\end{center}
\twocolumngrid
\setcounter{equation}{0}
\setcounter{figure}{0}
\setcounter{table}{0}
\setcounter{section}{0}
\setcounter{page}{1}
\makeatletter
\renewcommand{\theequation}{S\arabic{equation}}
\renewcommand{\thesection}{S-\Roman{section}}
\renewcommand{\thefigure}{S\arabic{figure}}
\renewcommand*{\thepage}{S\arabic{page}}
\renewcommand{\bibnumfmt}[1]{[S#1]}
\renewcommand{\citenumfont}[1]{S#1}

In this Supplemental Material, we provide arguments for the chosen friction function $\mu_A$ in Sec.~\ref{se:fric}, the dimensionless form of the equations in Sec.~\ref{se:dimless}, and details of our numerical solution method and convergence tests in Sec.~\ref{se:numerics}.

\section{The friction function}\label{se:fric}
The choice of friction function per species, $\mu_{(i)}$, crucially depends on the level of entanglement of the species~\cite{Sbrochard1983polymer,Sbrochard1991kinetics}. As we consider small B-mers, we assume that the B-mers are never entangled, allowing us to take $\mu_B\!=\!1-\phi$.

For the A-mers the situation is different. For the unentangled case, we need $\mu_A\!=\!\phi$, while for the entangled case we need $\mu_A\!=\!\alpha \phi$, where $\alpha\!=\!N_A/N_e\!>\!1$, and $N_e$ is a typical entanglement threshold. As we probe the dynamics valid for a wide range of $\phi$, we need to consider the proper case as a function of $\phi$.

\begin{figure}[h]
\includegraphics[width=0.46\textwidth]{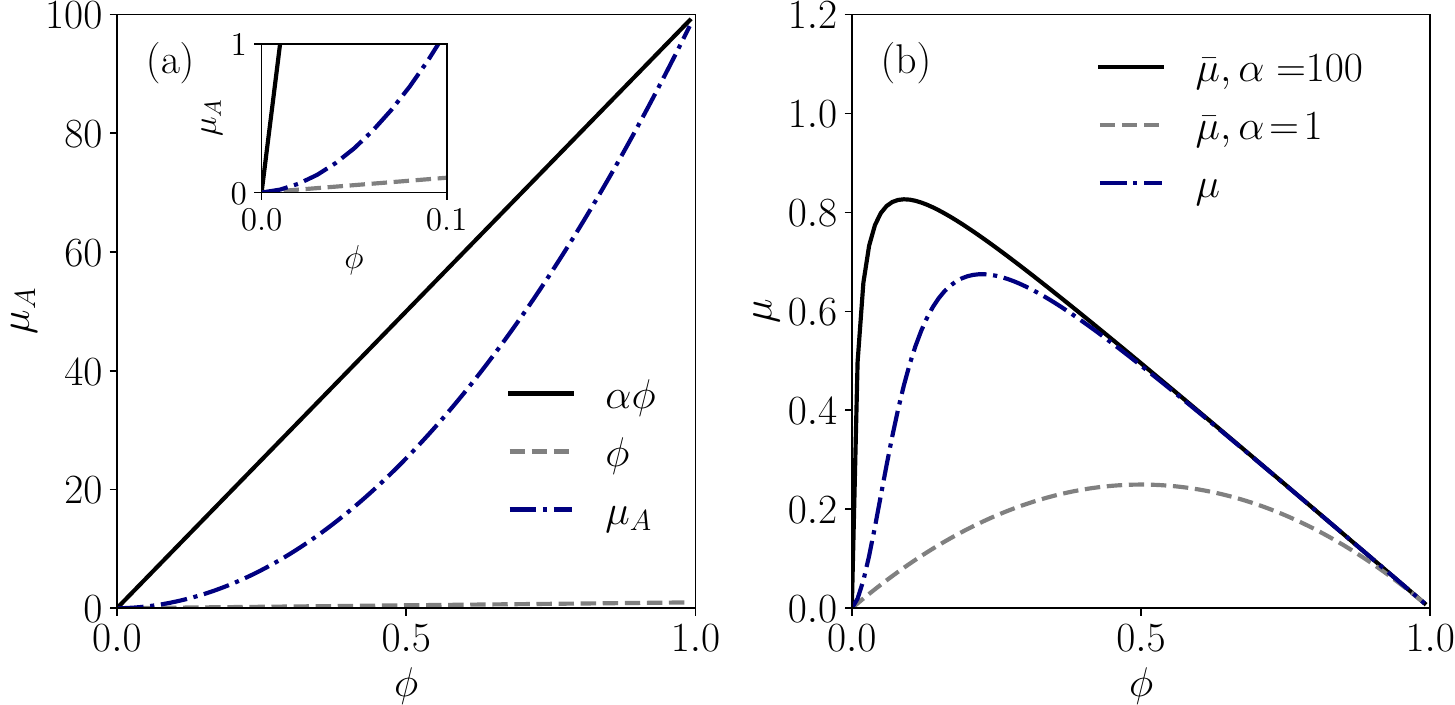}
\caption{A comparison of friction functions $\mu_A$ and $\mu$ for different entangled states (with $\alpha\!=\!100$). (a) A comparison between the canonical entangled friction $\alpha \phi$, the untangled friction $\phi$, and the $\mu_A$ used in our work. Inset: the $\phi\!\rightarrow\!0$ limit showing $\mu_A\!\rightarrow\!\phi$. (b) A comparison of the induced $\mu$ dependencies, using the canonical entangled friction $\mu_A\!=\!\alpha \phi$, the unentangled friction $\mu_A\!=\!\phi$, and the $\mu_A$ used in our work.}
\label{fig:fig1sm}
\end{figure}

We propose using $\mu_{A}\!=\!\phi\left[\alpha\phi+(1-\phi)\right]$. When $\phi\!\rightarrow\!0$, $\mu_A\rightarrow\!\phi$, which is the unentangled limit. As the local volume fraction increases, $\phi \!\rightarrow\! 1$, $\mu_A\!\rightarrow\! \alpha$, which is the entangled limit. We compare the suggested form of $\mu_A$ to the canonical entangled dependence, $\alpha \phi$, and the unentangled one $\phi$ in Fig.\ref{fig:fig1sm}(a).

Combined with $\mu_B\!=\!1-\phi$, we have
\begin{equation}\label{eq:mu}
    \mu = \phi\left(1-\phi\right)\frac{1-\phi+\alpha\phi}{1-\phi+\phi\left(1-\phi+\alpha\phi\right)} \ .
\end{equation}
We plot the above expression for $\mu$ compared with $\mu$ obtained using the untangled $\mu_A\!=\!\phi$, and the entangled $\mu_A\!=\!\alpha\phi$ (with $\mu_B\!=\!1-\phi$), which we denote as $\bar{\mu}$, in Fig.~\ref{fig:fig1sm}(b).

We now examine several limits of $\mu$ and the emerging dynamics. To ease the analysis, we note that for $\chi\!=\!0$ and $N_B\!=\!1$, $\bm{\Pi}\!=\!-\bm{I}\left[\log(1-\phi)+\left(1-N_A^{-1}\right)\phi\right]$, and $\Div \bm{\Pi}\!=\!\left[1+\phi(N_A-1)\right]\Grad\phi / \left[N_A(1-\phi)\right]$ (here we used the dimensionless $\bm{\Pi}$). First, in the low-dilution limit $\phi\!\ll\!1$, $\mu\!\rightarrow\!\phi$, and we recover a linear diffusive behavior $\frac{\phi(1-\phi)^2}{\mu}\Div\bm{\Pi}\rightarrow N_A^{-1} \Grad\phi$. Comparing this with a generic nonlinear diffusion term of the form $\phi^{m-1}\Grad \phi$ implies this case  corresponds to $m\!=\!1$. 

In the unentangled case for higher concentrations, $\alpha\!=\!1$, $\mu\!=\!\phi\left(1-\phi\right)$, and considering the long-polymer limit $N_A\!\rightarrow\!\infty$, $\frac{\phi(1-\phi)^2}{\mu}\Div\bm{\Pi}\rightarrow\phi \Grad\phi$, resulting in $m\!=\!2$. In the entangled limit ---  our main interest ---, we take $\alpha\!\rightarrow\!\infty$, leading to $\mu\!\rightarrow\!1-\phi$, implying $\frac{\phi(1-\phi)^2}{\mu}\Div\bm{\Pi}\rightarrow\phi^2 \Grad\phi$, corresponding to $m\!=\!3$ (see also~\cite{SLongPaper}). 

Another possible formalism~\cite{Sde1980dynamics,Spincus1981dynamics,Sbrochard1983polymer,Sbrochard1991kinetics} cast the dynamics of $\phi$ in terms of the flux $\bm{J}$
\begin{equation*}
    \partial_t \phi + \nabla\cdot\bm{J}\!=\!0 \ ,
\end{equation*}
where $\bm{J}\!=\!-D_M\left(\phi\right) \Grad \phi$, where $D_M$ is the mutual diffusion coefficient. It was suggested~\cite{Sbrochard1991kinetics} that for non-interacting ($\chi\!=\!0$) long and short chain mixtures $N_A\!\gg\!N_B$, the experimental results of~\cite{Sgreen1986matrix,Smills1984analysis,Sgreen1984limits,Sjordan1988mutual} suggest $D_M$ is
\begin{equation*}
    D_M\!=\!N_e \frac{\phi^2 (1-\phi)^2}{\zeta_0}\left[\frac{1}{N_A \phi} + \frac{1}{N_B(1-\phi)}\right]^2\ ,
\end{equation*}
which approaches $D_M\!\rightarrow\!\frac{N_e}{\zeta_0}\phi^2$ as $N_A$ increases (and $N_B\!\rightarrow\!1$), again corresponding to $m\!=\!3$, as claimed above.


\section{Dimensionless equations and numbers}\label{se:dimless}
In this section we systematically cast the equations of motion in dimensionless form. For convenience, in this section we use $\units{\bullet}$ notation to denote the units of each quantity and field denoting $L$ for lengths, $T$ for times, and $F$ for force, in $d$ spatial dimensions.

First, we note that $\units{\phi}\!=\!1$, $\units{\bm{u}}\!=\!LT^{-1}$, $\units{p}\!=\!\units{\bm{\Pi}}\!=\!\units{\bm{\sigma}}\!=\!FL^{-d+1}$. The continuity equation imply $\units{S}\!=\!T^{-1}$. Using $S\!=\!\delta^d(\bm{r}) Q t^{q-1}$ implies $\units{Q}\!=\!L^d T^{-q}$. The Flory-Huggins thermodynamic potential $F_{FH}\!=\!k_B T n_0 f(\phi)$ already uses $\units{k_BT}\!=\!FL$ (energy), and $\units{n_0}\!=\!L^{-d}$ (inverse volume). We also note that $\units{N_A}\!=\units{N_B}\!=\!1$, and $\units{\chi}\!=\!1$. The velocity difference equation necessitates $\units{n_0 \zeta_0 \mu\left(\bm{u}_A-\bm{u}_B\right)}\!=\!FL^{-d}$, implying $\units{\zeta_0}\!=\!F T L^{-1}$. Finally, the upper-convected Maxwell (UCM) model implies $\units{\lambda_{(i)}}\!=\!T$, and $\units{G_{(i)}}\!=\!FL^{-d+1}$. As $\units{\phi}\!=\!1$, $\units{G_A}\!=\!\units{G_0}$.

We now use $\lambda\!\equiv\!\lambda_A$ as our time scale (taking $\lambda_B\!\ll\!\lambda_A$), $n_0 k_B T$ to be our energy density unit, and $\sqrt{\frac{k_B T \lambda}{\zeta_0}}$ as our length-scale. As $\bm{\sigma}_B\!\simeq\!\eta_B\left(\bm{L}_B+\bm{L}_B^T\right)$, casting it in a unitless form yields a dimensionless number $\beta\!\equiv\!\eta_B / n_0 k_B T \lambda$, which is the ratio of the effective modulus $\eta_B/\lambda$ to the thermal energy density $n_0 k_B T$. Similarly, recasting the UCM relations yields the dimensionless number $G_0 / n_0 k_B T$ --- the ratio between the modulus $G_0$ and $n_0 k_B T$.

Overall, the dimensionless equations are then obtained as
\begin{gather*}
    \partial_t \phi_A + \nabla\cdot\left(\bm{u}_A \phi_A\right) = S \ , \\
    \partial_t \phi_B + \nabla\cdot\left(\bm{u}_B \phi_B\right) = 0 \ , \\
    \nabla\cdot\left(-p\bm{I}-\bm{\Pi} + \boldsymbol{\sigma}_{A} + \boldsymbol{\sigma}_{B}\right) = {\bm 0} \ , \\
    -\frac{1}{\mu}\left[\left(1-\phi\right)\nabla\cdot\left(\bm{\Pi}-\bm{\sigma}_{A}\right)+\phi\nabla\cdot\bm{\sigma}_{B}\right] = \bm{u}_A - \bm{u}_B \ , \\
    \UCD{\bm{\sigma}}_{A} = \tilde{G} \phi\left(\bm{L}_{A} + \bm{L}_{A}^T\right) - \bm{\sigma}_{A} \ ,
\end{gather*}
where all the fields and constants are dimensionless, and with $\bm{\sigma}_B\!=\!2\beta\left(\bm{L}_B+\bm{L}_B^T\right)$, and $\bm{\Pi}\!=\!\bm{I} \left(\phi\partial_\phi f-f\right)$.

Finally, we note in passing that the emerging effects of osmotic pressure divergence $\Div\bm{\Pi}$ are directly related to spatial gradients of the chemical potential $\phi \Grad\left(\partial_{\phi} f\right)$, as $\Div\bm{\Pi}\!=\!\Grad\left[\phi\partial_\phi f-f\right]\!=\phi \Grad \left(\partial_{\phi} f\right)$. We considered the formalism with $\bm{\Pi}$ to be consistent with the existing literature.

\section{Numerical simulation details}\label{se:numerics}

We utilized numerical simulations in both 1D and 2D~\cite{Svan1995python,Sharris2020array}. In both cases we adopted a ``staggered' approach, where velocity fields are defined on edges, and the concentration and stress fields at the centers of the discretized domain.

In 1D we chose a domain $x\in\left[-40,40\right]$, discretized to 2000 points. We chose the time-integration increment to be $dt\!=4\cdot10^{-4}$. After initializing all fields to zero, each time integration step consisted of evaluating the source (depending on its time dependence), solving for the average velocity field $v\!\equiv\!\phi u_A + (1-\phi) u_B$, and for the velocity difference $w\!\equiv\! u_A- u_B$, evaluating the velocity fields $u_A$ and $u_B$, and integrating in time $\phi$ and $\sigma$.

The source value was determined based on the amplitude $Q$, the power $q$, and the time $t$, as $\frac{Q t^{q}}{\sqrt{2\pi}}\exp{\left(-x^2/2\right)}$ (here we adopted a typical unity width). Adding the two continuity equations yielded $\partial_x v = S$. We solved this equation by simply integrating $S$ using cumulative trapezoid integration from the origin.

We then solve for $w$ using $\Pi$ and $\sigma_A$ evaluated at the centers of the grid, and employing a first-order spatial derivative to transform the derivative to the edges. We also had to evaluate the mobility on the edges --- we did this by a simple nearest-neighbor averaging of $\phi$ from the two neighboring centers. Finally, we constructed the operator of the difference equation, and using a sparse solver to obtain the solution for $w$. Once we had $v$ and $w$, we evaluated $u_A\!=\!v+(1-\phi)w$ and $u_B\!=\!v-\phi w$.

\begin{figure}[b!]
\includegraphics[width=0.48\textwidth]{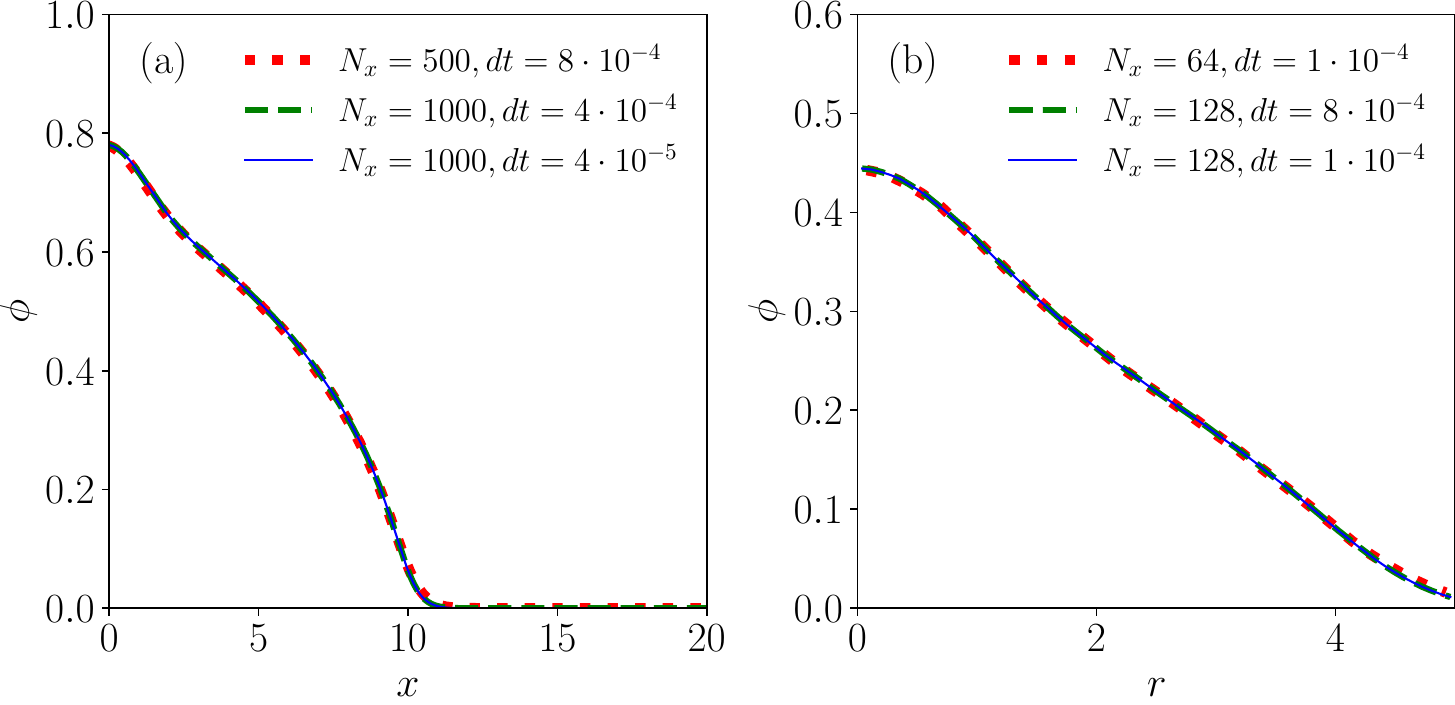}
\caption{Convergence tests for the 1D simulations in (a) and for the 2D simulations in (b). (a) $\phi$ profiles obtained from 1D simulations (with $N_A\!=\!10^4$, $N_B\!=\!1$, $\alpha\!=\!100$, $\beta\!=\!1$, $\tilde{G}\!=\!1$, and $Q\!=\!0.1$) integrated to $t\!=\!99$ (in dimensionless time), showing similar concentration profiles for different $dt$ and $dx$ values, as indicated by the legends. (b) Radially-averaged $\phi$ profiles from 2D simulations (with similar parameters as the 1D simulations) integrated to $t\!=\!114.9$ showing convergence for different $dt$ and $dx$ values.}
\label{fig:fig2sm}
\end{figure}

To integrate $\phi$ temporally, we evaluated the flux $u_p \bar{\phi}$, where $\bar{\phi}$ indicates $\phi$ evaluated on the edges (using a simple averaging, as described above), and used a first-order upwind scheme to integrate $\phi$ by $dt$. We used a similar upwind scheme to integrate the stress $\sigma_A$.

Our 2D simulations used a grid of $128\times128$, spanning $\left(x,y\right)\in\left[-5,5\right]^2$, and a $dt\!=\!10^{-4}$. We used here a similar staggered-grid approach, and a similar integration step algorithm. Due to the large size of our system, instead of using a sparse linear solver, we utilized a conjugate gradient solver, improving our running times. 

We show a convergence test for both the 1D and 2D simulations in Fig.~\ref{fig:fig2sm}. For a fixed set of physical parameters ($N_A\!=\!10^4$, $\alpha\!=\!100$, $\beta\!=\!1$, $\tilde{G}\!=\!1$, $Q\!=\!0.1$ with $q\!=\!1$), we vary the spatial resolutions $dx$s and the time-integration increments $dt$s. The results of Fig.~\ref{fig:fig2sm} indicate that our simulations are well-converged for a wide range of spatial and temporal discretizations.

\end{document}